\documentclass{emulateapj}
\usepackage{apjfonts}
\begin{document}
\title{On Kinematic Substructure in the Sextans Dwarf Spheroidal Galaxy\footnote{Based on observations using the Magellan Telescopes}}
\shorttitle{Kinematic Substructure in Sextans}
\author{Matthew G. Walker\altaffilmark{1}, Mario Mateo\altaffilmark{1}, Edward W. Olszewski\altaffilmark{2}, Jayanta Kumar Pal\altaffilmark{3}, Bodhisattva Sen\altaffilmark{3}, and Michael Woodroofe\altaffilmark{3}}
\altaffiltext{1}{Department of Astronomy, University of Michigan, 830 Dennison Building, Ann Arbor, MI 48109-1042}
\altaffiltext{2}{Steward Observatory, The University of Arizona, Tucson, AZ 85721}
\altaffiltext{3}{Department of Statistics, University of Michigan, 439 West Hall, Ann Arbor, MI 48109-1107}
\begin{abstract} 
We present multifiber echelle radial velocity results for 551 stars in the Sextans dwarf spheroidal galaxy and identify 294 stars as probable Sextans members.  The projected velocity dispersion profile of the binned data remains flat to a maximum angular radius of $30^{\prime}$.  We introduce a nonparametric technique for estimating the projected velocity dispersion surface, and use this to search for kinematic substructure.  Our data do not confirm previous reports of a kinematically distinct stellar population at the Sextans center.  Instead we detect a region near the Sextans core radius that is kinematically colder than the overall Sextans sample with $95\%$ confidence.
\end{abstract}
\keywords{galaxies: dwarf --- galaxies: individual (Sextans) --- galaxies: kinematics and dynamics --- (galaxies:) Local Group}
\section{Introduction, Observations and Data}
\label{sec:intro}

The most recent kinematic studies of dwarf spheroidal (dSph) galaxies show that radial velocity (RV) dispersion profiles of these dark matter dominated systems are generally flat, with some evidence for a declining dispersion at large radius in Draco and Ursa Minor (Wilkinson et al.\ 2004, but see Mu\~{n}oz et al.\ 2005; Walker et al.\ 2006; Mateo et al.\ {\em in prep.}).  However, localized regions having enhanced stellar density and/or comparatively ``cold'' kinematics have been detected in some dSphs (Olszewski \& Aaronson 1985; Kleyna et al.\ 2003; Kleyna et al.\ 2004, ``K04'' hereafter; Coleman et al.\ 2004).  Such substructure may be related to the star formation history of individual systems (Tolstoy et al.\ 2004; Olszewski et al.\ 2006), or may hint at a merger history (Coleman et al.\ 2004; Wilkinson et al.\ 2005).  The presence of substructure in these tiny galaxies has cosmological implications as well.  Kleyna et al.\ (2003) demonstrate that the apparent substructure in Ursa Minor should quickly disperse in the presence of a centrally cusped dark matter halo, but can remain intact for a Hubble time in the potential of a cored halo.  This raises doubts about whether the smallest galaxies conform to the universal density profile that results from simulations of cold dark matter (Navarro, Frenk \& White, 1997).

Here we present first results from an ongoing survey of dSph stellar RVs.  We used the Michigan-MIKE Fiber System (MMFS) at the 6.5m Magellan/Clay telescope at Las Campanas Observatory, in March 2004 and February 2005 to obtain spectra of target stars selected from V,I photometry of the Sextans red giant branch.  A complete description of MMFS is forthcoming (Walker et al.\ {\em in prep.}).  Briefly, 256 fibers placed over a field of diameter $30^{\prime}$ feed the two channels of the MIKE echelle spectrograph (Bernstein et al. 2003).  Order-blocking filters isolate the spectral region 5140-5180 \AA, which contains strong Mg-I triplet absorption features and is free of the sky emission lines that contaminate the infrared Ca triplet.  After binning, the spectra have resolution $R \sim 20000$ (blue channel) and $R \sim 15000$ (red channel).

The data reduction procedure is similar to that described in Mateo et al.\ (2006, {\em in prep.}).  Briefly, after extracting and calibrating spectra, we measure RV using the IRAF task FXCOR, cross-correlating the spectrum of each target star against that of a high-S/N template composed of co-added spectra of RV standard stars.  A series of quality controls yields a data set containing 810 RV measurements for 551 Sextans candidate members.  We use multiple measurements of 173 stars from 12 partially-overlapping fields to estimate measurement uncertainties  as in Walker et al.\ (2006).  The mean $1\sigma$ RV uncertainty is $\pm2.6$ km s$^{-1}$.  

\begin{figure}
\figurenum{1}
\label{fig:velocities}
\plotone{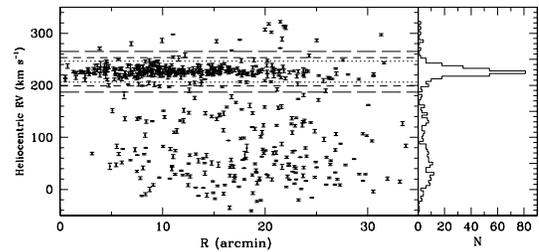}
\caption{{\em left:} Heliocentric RV vs.\ angular distance from the Sextans center ($\alpha_{2000}=$10:13:03, $\delta_{2000}=$-01:36:54; Mateo 1998).  Dotted, dashed, and long-dash lines mark boundaries of N=276, N=294, and N=303 member samples, respectively.  Not shown are 21 observed stars at $60^{\prime} \leq R \leq 80^{\prime}$, none of which are likely members.  {\em right:} Histogram of the Sextans RV distribution.}
\end{figure}

Figure \ref{fig:velocities} plots the measured heliocentric RV versus angular distance from the Sextans center.  The broad distribution of RV outliers away from the main peak at 226 km s$^{-1}$ is consistent with the RV distribution of foreground Milky Way dwarfs (whose magnitudes and colors place them inside our Sextans RGB selection region) predicted by the Besancon Milky Way model (Robin et al.\ 2003)\footnote{see http://bison.obs-besancon.fr/modele}.  To determine membership we use a biweight estimator (Beers et al.\ 1990) of distribution center and variance, and iteratively reject RVs that lie more than $3\sigma$ from the distribution center.  This algorithm yields a default sample of 294 probable Sextans members.  Because the conventional $3\sigma$ cutoff is somewhat arbitrary, we repeat our analyses for possible samples containing 276 stars ($2.6\sigma$ outlier rejection), and 303 stars ($4\sigma$).  

\section{Radial Velocity Dispersion: Profile and Surface}
\label{sec:profile}

\begin{figure}
\figurenum{2}
\label{fig:vmap}
\plotone{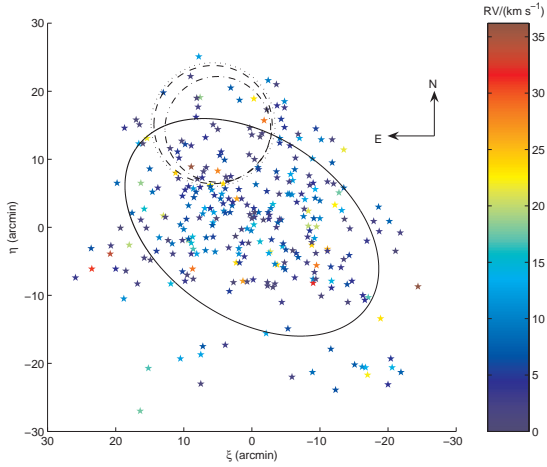}
\caption{Sky map, in standard coordinates, of 303 probable Sextans member stars with velocities measured using MMFS.  Color scale indicates the magnitude of each stellar RV relative to the sample mean.  The ellipse has geometric mean radius corresponding to $r_{core}=16.6^{\prime}\pm1.2^{\prime}$ (Irwin \& Hatzidimitriou 1995).  The dashed (dotted, dot-dash) circle encloses the region from the N=294 (N=276, N=303) sample where the measured RVs are most likely to have been drawn from two distinct Gaussian distributions.}
\end{figure}

Figure \ref{fig:vmap} maps the projected positions of the likely Sextans members and indicates the magnitude of each star's RV relative to the sample mean.  We detect no rotation---the maximum rotation signal of $0.9$ km s$^{-1}$ is exceeded in $43\%$ of Monte Carlo trials.  Following the procedure described in Walker et al.\ (2006), we estimate from the default N=294 sample a mean RV of $225.8 \pm 0.5$ km s$^{-1}$ and RV dispersion $8.9 \pm 0.4$ km s$^{-1}$, in agreement with K04.  For N=276 and N=303 samples the RV dispersion is $7.1 \pm 0.3$ and $10.3 \pm 0.5$ km s$^{-1}$, respectively.  Binning by radius, we estimate the projected RV dispersion profile, $\sigma_p(R)$, using the Gaussian maximum-likelihood method of K04 (see their Equation 1).  The RV dispersion calculated for a given bin is not affected if we modify K04's Equation 1 to include a fractional component of Galactic contaminants with plausible (e.g., uniform within the accepted range) RV distribution. 

\begin{figure}
\figurenum{3}
\label{fig:profiles}
\plotone{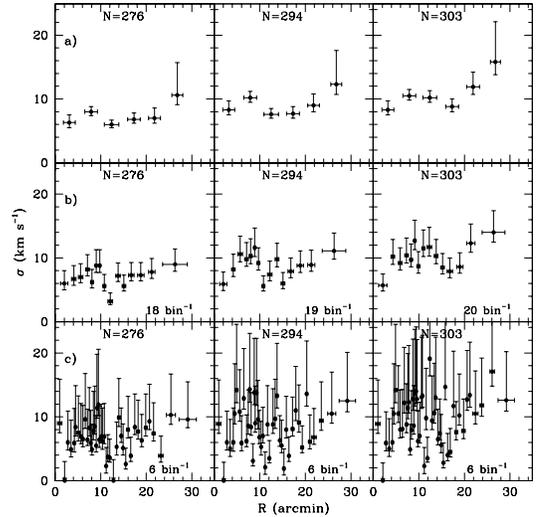}
\caption{Sextans projected velocity dispersion profile measured from the N=276, N=294, and N=303 member samples.  Bins in (a) have regularly-spaced outer boundaries.  In b-c, bins contain approximately equal numbers of stars.  Horizontal error bars indicate the standard deviation of R values.}    
\end{figure}

Figure \ref{fig:profiles} plots the Sextans RV dispersion profile calculated using a variety of bin sizes.  Aside from setting the global RV dispersion value, choice of membership sample has little effect on the measured profile.  As in K04, the top panels of Figure \ref{fig:profiles} use bins having regularly-spaced outer boundaries at $5^{\prime},10^{\prime},...,30^{\prime}$.  Our data do not reach the largest radii sampled by the K04 data, so we can neither confirm nor rule out the falling dispersion K04 detect at $R \geq 30^{\prime}$.  For $R < 30^{\prime}$ our data indicate a flat Sextans profile.  In contrast, K04 measured $\sigma_P \leq 1$ km s$^{-1}$ for the 7 stars in their sample at $R < 5^{\prime}$ (see Figure 4 of K04).  The 40 stars from our sample that occupy this region have $\sigma_p=8.3_{-0.8}^{+1.4}$ km s$^{-1}$.  The profiles shown in Figure \ref{fig:profiles}($b,c$) have annuli chosen such that bins contain equal numbers of stars.  A ``cold'' inner profile point begins to emerge as the number of bins increases, but at the cost of diminished statistical significance.  The profiles in ($c$) use 43 bins (6 stars per bin), the smallest number for which we find any annulus with $\sigma_p \leq 1$ km s$^{-1}$.  Monte Carlo simulations indicate that one expects to measure such a small dispersion in $\sim 1$ of 43 bins even if the true profile is perfectly flat.\footnote{It is perhaps intriguing that our six stars contributing to the cold inner point in Figure \ref{fig:profiles}c are localized---all lie within $3^{\prime}$ of the Sextans center and five occupy a region of diameter $2^{\prime}$.  However, Gaussian-random samples of 294 stars contain, on average, 3 mutually exclusive sets of six neighboring stars for which we would measure $\sigma_p \leq 1$ km s$^{-1}$.  Including the stars contributing to the cold point in Figure \ref{fig:profiles}c, we detect two such sets in our sample.  As they are consistent with what is expected to arise by chance, we conclude that these detections alone do not provide compelling evidence of cold substructure.}

We conclude that our data are consistent with a flat Sextans RV dispersion profile for $R < 30^{\prime}$.  This result is consistent with K04 if one omits their innermost point, as K04 do when using their observed profile to estimate a Sextans mass of M$(r \leq 1$ kpc$) \sim 10^7 - 10^8$ M$_{\sun}$.  We require a more radially extended data set in order to further constrain the Sextans mass.

Our data set consists of measurements $\{(X_1,Y_1,V_1,\sigma_1),...,(X_N,Y_N,V_N,\sigma_N)\}$, where $X_i$ and $Y_i$ give the projected position of the $i^{th}$ member star with respect to the Sextans center, and $V_i$ and $\sigma_i$ are the measured RV and its formal uncertainty.  To identify potential kinematic substructure we search for regional anomalies in the RV distribution as a function of projected position.  From the data we estimate the Sextans RV dispersion {\em surface} using a non-parametric Nadaraya-Watson (1964) estimator: 
\begin{equation}
  \label{eq:surface}
  \hat{\langle v^2 \rangle} (x,y)=\frac{\sum_{i=1}^{N}\bigl[(V_i-\langle V \rangle)^2-\sigma_i^2\bigr]K(\frac{x-X_i}{h_x},\frac{y-Y_i}{h_y})}{\sum_{i=1}^NK(\frac{x-X_i}{h_x},\frac{y-Y_i}{h_y})},
\end{equation}
where $\langle V \rangle$ is the sample mean, $K(\frac{x-X_i}{h_x},\frac{y-Y_i}{h_y})$ is a smoothing kernel and $h_x$, $h_y$ specify spatial smoothing bandwidths.  This estimator provides a weighted local average of $\hat{\langle v^2 \rangle}$ on a grid of $(x,y)$.  For simplicity we adopt a bivariate Gaussian kernel with isotropic smoothing: $K(\frac{x-X_i}{h_x},\frac{y-Y_i}{h_y}) \propto \exp[-\frac{1}{2}\frac{(x-X_i)^2+(y-Y_i)^2}{h_s^2}]$, where $h_s=h_x=h_y$.  Since the data points have a nonuniform spatial distribution, we allow the spatial smoothing bandwidth, $h_s$, to vary over the grid surface.  At every $(x,y)$, $h_s$ takes the minimum value for which there exist $n$ data points located within $3h_s$ of $(x,y)$.  The value chosen for $n$ determines the number of neighboring data points that contribute significantly to the kernel.  

\begin{figure*}
\figurenum{4}
\label{fig:contour294}
\plotone{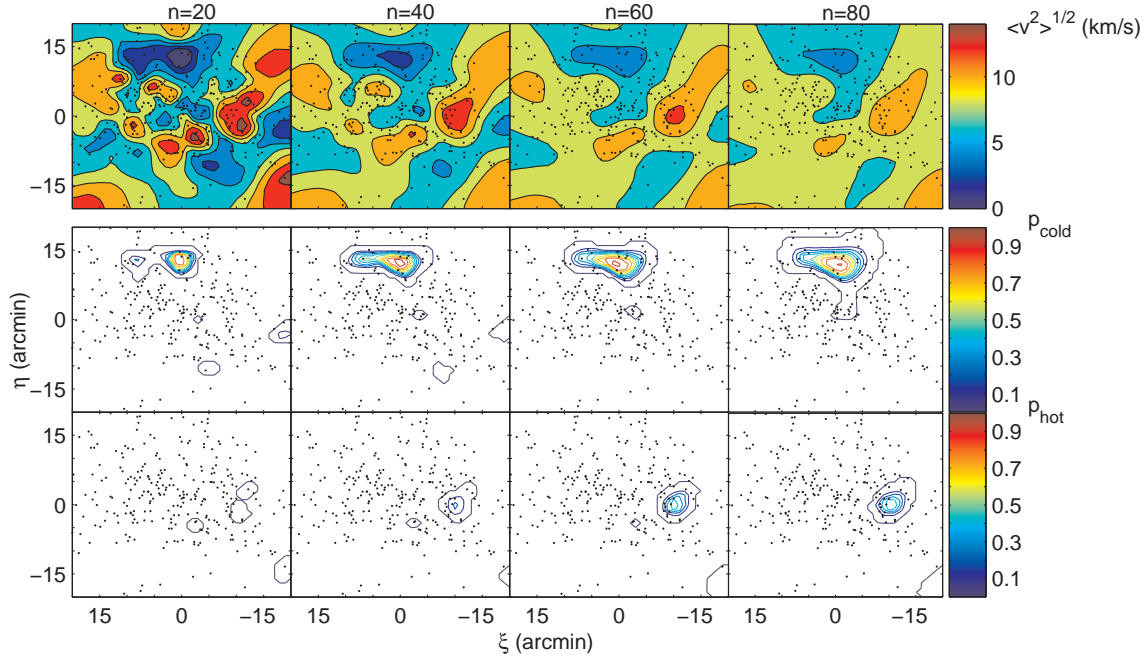}
\caption{{\em Top row}: Contours of the Sextans radial velocity dispersion surface estimate, $\hat{\langle v^2 \rangle}^{1/2}(x,y)$, obtained by applying Equation \ref{eq:surface} to the N=294 member sample.  From left to right, panels indicate results from estimations using nearest-neighbor parameter $n=20,40,60,80$.  Black markers indicate the positions of member stars used in the estimation.  {\em Middle}/{\em Bottom row:} Contours of statistical significance for regions of apparent cold/hot kinematic substructure.  $p_{cold}(x,y)$ is the fraction of artificial data sets for which the simulated surface minimum is greater than the estimate from the Sextans data at $(x,y)$.  $p_{hot}(x,y)$ is the fraction of artificial data sets for which the simulated surface maximum is less than the estimate from the Sextans data at $(x,y)$.  Advancing from outermost to innermost, contours enclose regions where $p(x,y) > 0.0,0.1,0.2,...,1.0$.}
\end{figure*}

We use Equation \ref{eq:surface} to estimate the Sextans RV dispersion surface, $\hat{\langle v^2 \rangle}^{1/2}_{S}(x,y)$, repeating for neighbor values $n=10,20,...,150$.  In all cases we evaluate Equation \ref{eq:surface} on a square grid with $x$ and $y$ values each spanning the range $\{-20^{\prime},-19^{\prime},...,+19^{\prime},+20^{\prime}\}$.  The top row of panels in Figure \ref{fig:contour294} maps contours of the Sextans RV dispersion surface resulting from a subset of these estimations with N=294 and $n=20,40,60,80$.  

To assess the significance of apparent surface features we adopt as a null hypothesis that the measured RVs are drawn everywhere from the same (not necessarily Gaussian) distribution.  
To simulate the null hypothesis we generated (for each of the three Sextans samples) 1000 artificial data sets in which the stars have identical positions to those in the Sextans member sample, but in which $(V_i,\sigma_i)$ pairs are drawn randomly, without replacement, directly from the Sextans data.  All the observed Sextans RVs are present in each artificial data set, but scrambled with respect to position.  This dissociates any existing correlation between position and RV while preserving the global RV distribution. 

We apply Equation \ref{eq:surface} to each artificial data set and record the distributions of artificial surface minima and maxima, $\hat{\langle v^2 \rangle}_{A,min}$ and $\hat{\langle v^2 \rangle}_{A,max}$.  The significance of a small Sextans surface dispersion at grid point $(x,y)$ is then given by the fraction, $p_{cold}(x,y)$, of artificial surfaces for which $\hat{\langle v^2 \rangle}_{S}(x,y) < \hat{\langle v^2 \rangle}_{A,min}$; conversely, the significance of a large Sextans dispersion at $(x,y)$ is given by the fraction, $p_{hot}(x,y)$, of artificial surfaces for which $\hat{\langle v^2 \rangle}_{S}(x,y) > \hat{\langle v^2 \rangle}_{A,max}$.  The bottom two rows of panels in Figure \ref{fig:contour294} map contours of $p_{cold}(x,y)$ and $p_{hot}(x,y)$ corresponding to each Sextans RV surface estimation.

For nearly all $n$ and all three cases of membership, the Sextans RV dispersion surface has a minimum value near a point $\sim 15^{\prime}$ north of the Sextans center.  For the default N=294 member sample the significance is $p_{cold} > 0.9$ for estimations in which $30 \leq n \leq 80$, and reaches $p_{cold}=0.965$ for $n=40$.  The same northern region remains the coldest in estimations using the N=276 and N=303 member samples, despite the fact that the former sample removes 2 stars from within $10^{\prime}$ of the region's center, and the latter contributes 3 high-velocity stars to the region.  For N=276, the significance is $p_{cold} > 0.9$ for a wide range of $n$ and reaches $p_{cold}=0.948$ for $n=60$.\footnote{The N=276 Sextans sample contains a second local minimum, $10^{\prime}$ south of the Sextans center, that is significant with $p_{cold} \sim 0.95$ in estimations using $20 \leq n \leq 40$.  Unlike the northern feature, however, this feature disappears when considering less restrictive membership cases.}  For N=303 the northern feature remains the Sextans surface minimum, but with only $p_{cold} \leq 0.671$.  The RV distribution at the Sextans {\em{center}} does not provide compelling evidence against the null hypothesis.  For N=276 and N=294, all surface points inside $R < 5^{\prime}$ have $p_{cold} < 0.25$; for N=303 the center has $p_{cold} < 0.5$, for all $n$. 

We have also employed the parametric test described by Kleyna et al.\ (2003; K04) which asks whether subregions of the position-RV data are more likely described by a single Gaussian or by a mixture of two distinct Gaussians.  This test yields results similar to those from our nonparametric test.  For the N=294 sample, a model in which RVs from $80\%$ of the 35 stars centered on a location $15^{\prime}$ NNE of the Sextans center (dashed circle in Figure \ref{fig:vmap}) are drawn from a secondary Gaussian with disperion $2.7$ km s$^{-1}$ has likelihood $3.7 \times 10^4$ times larger than a global Gaussian model.  Simulations using artificial data sets indicate this result is sufficient to rule out our null hypothesis with significance $p=0.968$ (here, $p$ indicates the fraction of artificial data sets for which no subregion yields as large a likelihood ratio).  For N=276, the northern region is kinematically cold ($80\%$ of 35 stars drawn from a secondary Gaussian with 2.8 km s$^{-1}$ dispersion) with significance $p=0.944$.  For N=303, the region remains the strongest candidate for cold substructure ($60\%$ of 35 stars drawn from a secondary Gaussian with $0$ km s$^{-1}$ dispersion), but the significance falls to $p=0.756$.  No aperture overlapping the Sextans center favors a mixed Gaussian model with $p > 0.5$, for any of the three samples and a variety of aperture sizes.  

The striking agreement between parametric and nonparametric tests prompts two conclusions: 1) we do not confirm the presence of a kinematically cold Sextans core, and 2) we find stronger evidence ($p \sim 0.95$ for N=276 and N=294) for cold substructure north of center, near the Sextans core radius.  The latter result is somewhat sensitive to membership selection---when the most marginal members are allowed into the sample, the significance of the feature drops markedly.

\section{Discussion}
\label{sec:discussion}
Kleyna et al.\ have graciously provided us with their RV data for 118 Sextans stars (88 members).  Velocity measurements for the 40 stars common to our survey generally agree to within formal uncertainties.  For $R < 5^{\prime}$, the data sets contain 2 stars in common, with excellent RV agreement.  What accounts for the discrepant results between the two studies?  Sample size must play a role.  K04 propose the existence of a kinematically distinct Sextans core based on the cold RV dispersion of the seven innermost stars in their sample.  We find that if we restrict our analyses to similar numbers of stars, we obtain similar evidence.  Our RV dispersion profile (Figure \ref{fig:profiles}c) identifies a group of six stars near the Sextans center that have $\sigma_p=0.0_{+0.4}^{+3.3}$ km s$^{-1}$.  If we apply the Gaussian mixture test to the five of these stars clustered nearest a point $3^{\prime}$ north of the Sextans center, a scenario in which four are drawn from a secondary Gaussian with $0$ km s$^{-1}$ dispersion is 115 times more likely than the alternative in which all five are drawn from the main Sextans distribution (for comparison, K04 calculate a likelihood ratio of $195$ for their cold inner sample).  At the location of these five stars, the squared RV dispersion estimated using Equation \ref{eq:surface} with $n=5$ is negative, indicating that the measurement uncertainties dominate the scatter among the local velocities.  However, for our data neither test result is significant at more than the $p=0.5$ level.  

The substructure candidate we detect lies near the Sextans core radius.  If we assume $80\%$ (the fraction indicated by the Gaussian mixture test) of the stars in the $15^{\prime}$ spanned by this feature belong to a distinct ``cluster'' of stars, and adopt a surface brightness over this region that is constant and equal to one-half Sextans' central surface brightness, we estimate a ``cluster'' luminosity of $3 \times 10^4$ L$_{\sun}$.  This is similar to the luminosity Kleyna et al.\ (2003) ascribe to the cold clump in UMi, but an order of magnitude smaller than the luminosity K04 estimate for a nominal core cluster in Sextans.  The timescale for the orbital decay of our ``cluster'' due to dynamical friction would therefore be an order of magnitude larger than the 0.7-1.5 Gyr K04 calculate for a core cluster (assuming the same cluster M/L and Sextans halo parameters).  This would remove the immediate difficulty in explaining why, if the substructure we detect is associated with a bound cluster, it has not spiralled to the Sextans center.  If such a cluster exists it should be detectable with deep imaging.

Goerdt et al.\ (2006) provide an explanation for the presence of substructure near the core radius of a dSph.  They simulate the evolution of globular cluster orbits in the Fornax dSph for both cored and cusped inner dark matter halos.  For a cusped halo, dynamical friction drags clusters to the Fornax center on timescales less than 5 Gyr.  For a cored halo, the infalling clusters ``stall'' indefinitely near the dSph core radius, a consequence of what Goerdt et al.\ interpret as a type of orbital resonance.  Thus the presence of five globular clusters near Fornax's core radius strongly favors a cored dark matter halo model.  In this context it is intriguing that Ursa Minor and perhaps Sextans also show signs of kinematically cold substructure near their core radii.  

We are grateful to Steve Shectman, Steve Gunnels, Patrick Kuschack and Alex Athey for assistance in developing MMFS, and to the staff at Las Campanas Observatory for supporting MMFS.  We thank the anonymous referee for helpful comments.  This work is supported by NSF grants AST 05-07453, AST 02-06081, AST 00-98518, AST 02-05790, and AST 05-07511, and by the Horace H. Rackham School of Graduate Studies at the University of Michigan.

\end{document}